# Integration of Simultaneous Resting-State Electroencephalography, Functional Magnetic Resonance Imaging, and Eye-Tracker Methods to Determine and Verify Electroencephalography Vigilance Measure

Ahmad Mayeli[1,2], Obada Al Zoubi[1,2], Masaya Misaki[1], Jennifer L. Stewart[1], Vadim Zotev[1], Qingfei Luo[1], Raquel Phillips[1],

Stefan Fischer[3], Marcus Götz[3], Martin P. Paulus[1], Hazem Refai[2], and Jerzy Bodurka[1,4]

[1]Laureate Institute for Brain Research, Tulsa, OK, United States

[2]Electrical and Computer Engineering, University of Oklahoma, Tulsa, OK, United States

[3]MRC Systems GmbH, Heidelberg, Germany

[4]Stephenson School of Biomedical Engineering, University of Oklahoma, Tulsa, OK, United States

Address correspondence to:
Jerzy Bodurka
Laureate Institute for Brain Research
6655 South Yale Avenue, Tulsa, OK 74136
E-mail: jbodurka@laureateinstitute.org
Phone: +1-918-502-5101

Number of figures: 6

Number of tables: 3

# Abstract

*Background/Introduction:* Concurrent electroencephalography and resting-state functional magnetic resonance imaging (rsfMRI) have been widely used for studying the (presumably) awake and alert human brain with high temporal/spatial resolution. Although rsfMRI scans are typically collected while individuals are instructed to focus their eyes on a fixated cross, objective and verified experimental measures to quantify degree of vigilance are not readily available. Electroencephalography (EEG) is the modality extensively used for estimating vigi- lance, especially during eyes-closed resting state. However, pupil size measured using an eye-tracker device could provide an indirect index of vigilance.

*Methods:* Three 12-min resting scans (eyes open, fixating on the cross) were collected from 10 healthy control participants. We simultaneously collected EEG, fMRI, physiological, and eye-tracker data and investigated the correlation between EEG features, pupil size, and heart rate. Furthermore, we used pupil size and EEG features as regressors to find their correlations with blood-oxygen-level-dependent fMRI measures.

*Results:* EEG frontal and occipital beta power (FOBP) correlates with pupil size changes, an indirect index for locus coeruleus activity implicated in vigilance regulation ($r = 0.306$, $p < 0.001$). Moreover, FOBP also correlated with heart rate ($r = 0.255$, $p < 0.001$), as well as several brain regions in the anticorrelated network, including the bilateral insula and inferior parietal lobule.

*Discussion:* In this study, we investigated whether simultaneous EEG-fMRI combined with eye-tracker measurements can be used to determine EEG signal feature associated with vigilance measures during eyes-open rsfMRI. Our results support the conclusion that FOBP is an objective measure of vigilance in healthy human subjects.

Impact Statement

We revealed an association between electroencephalography frontal and occipital beta power (FOBP) and pupil size changes during an eyes-open resting state, which supports the conclusion that FOBP could serve as an objective measure of vigilance in healthy human subjects. The results were validated by using simultaneously recorded heart rate and functional magnetic resonance imaging (fMRI). Interestingly, independently verified heart rate changes can also provide an easy-to-determine measure of vigilance during resting-state fMRI. These findings have important implications for an analysis and interpretation of dynamic resting-state fMRI connectivity studies in health and disease.

# 1. Introduction

Resting-state functional magnetic resonance imaging (rsfMRI) has become an important tool for studying the human brain due to its simplicity and non-invasiveness, as well as its requisite of least effort from the subjects (Fox and Greicius, 2010; Lee et al., 2013; Smitha et al., 2017). rsfMRI scans are conducted with eyes-closed, eyes-open, or eyes fixating on a cross. Cross fixation is the most often used paradigm and produces the most reliable results (Patriat et al., 2013; Yuan et al., 2013). However, it is unclear how well individuals manage to keep their eyes open and their gaze on the fixated cross over time, using this paradigm. In addition, an individual's degree of changes in vigilance during rsfMRI may affect functional connectivity results in both the cortical and subcortical brain regions (Falahpour et al., 2018). Therefore, to obtain a measure of vigilance during rsfMRI scans, employing another modality, such as eye-tracking or electroencephalography (EEG), is necessary. Notable, vigilance is a term which has been used with various definitions by different groups of scientists (Oken et al., 2006; Sander et al., 2015), such as ability to sustain attention to a task for a period of time (Davies and Parasuraman, 1982; Foxe et al., 2012) and attention to potential threats or dangers (Mogg et al., 2000). One of the most common definitions for vigilance is tonic alertness (Chang et al., 2016; Posner, 2008; Schneider et al., 2016), which has been used in this study.

Independent and concurrent modality signals such as EEG and eye-tracker could be used to continuously characterize one's fluctuations in vigilance, allowing researchers to use this vigilance index to assess fMRI data quality better. Even though the eye-tracker setup might be easier than EEG during fMRI acquisition, EEG is the modality that widely used simultaneously with fMRI. The reason for that is, simultaneous EEG-fMRI leverages fMRI's capacity to measure whole brain hemodynamic activities at the high spatial resolution of fMRI and high temporal resolution of EEG signals, directly reflecting electrophysiological brain activities (Niazy et al., 2005). In other words, unlike eye-tracker, EEG recording during fMRI acquisition could provide additional brain information except for subject's vigilance level.

Indeed, EEG is the modality most extensively used for monitoring vigilance, specifically during eyes-closed resting states (Berry et al., 2012; Hegerl et al., 2008; Horovitz et al., 2008; Olbrich et al., 2009; Sadaghiani et al., 2010; Wong et al., 2013). Slow-wave EEG is mostly studied during sleep (Ferrarelli et al., 2019; Korf et al., 2017), since delta (<4 Hz) and theta (4–7 Hz) waves dominate the EEG signal during drowsiness and sleep (Korf et al., 2017). In contrast, alpha (7-13 Hz) and beta (13-30 Hz) rhythms and the ratio of high to low-frequency band power have been employed to measure vigilance levels during resting states as well as during active tasks (Belyavin and Wright, 1987; Chang et al., 2016; Makeig and Inlow, 1993; Sadaghiani et al., 2010). For instance, (Hegerl et al., 2008) defined three vigilance states based on EEG frequency characteristics during an eyes-closed rsfMRI recording: (1) alertness and relaxed wakefulness, characterized by dominant alpha activity, (2) drowsiness, classified by dissolving alpha activity and theta rhythm dominance; and (3) sleep, marked by slow-wave activity and sleep spindles. Further, they validated the results of the EEG vigilance classification by analyzing the heart rates during the different brain states (Olbrich et al., 2009). When these EEG vigilance states were correlated with simultaneously recorded rsfMRI data, findings demonstrated that decreased vigilance was linked to a higher blood-oxygen-level-dependent (BOLD) signal in the parietal/occipital cortices as well as in regions of the prefrontal and anterior cingulate cortices (PFC and ACC); in addition, decreased vigilance was associated with reduced BOLD signal within the thalamus as well as in other specific regions of PFC (Olbrich et al., 2009). An additional eyes-closed rsfMRI study recording simultaneous EEG (Sadaghiani et al., 2010) introduced global field power (GFP) of upper alpha band (10–12 Hz) oscillations as the most consistent EEG index of tonic alertness; moreover, the GFP time course of upper alpha band positively correlated with BOLD signal fluctuations within the dorsal ACC, anterior insula, anterior PFC, and thalamus. A recent study defined an EEG wakefulness index, named as EWI, relied on both spectral and topographical instantaneous EEG information during eyes-closed resting state (Knaut et al., 2019). Given that eyes-closed and eyes-open conditions provide divergent EEG measures of vigilance with respect to brain topography and power (Barry et al., 2007) as well as differing BOLD signal patterns (Patriat et al., 2013; Zou et al., 2009; Zou et al., 2015a; Zou et al., 2015b), EEG vigilance measures used for eyes-closed resting states may not be easily applied to eyes-open paradigms. More specifically, while several studies showed the higher EEG alpha power is associated with higher vigilance level during eyes closed, it is noted in (Oken et al., 2006), during eyes open resting state, higher EEG alpha power may be associated with lower alertness.

Another potential solution for monitoring vigilance during eyes-open rsfMRI is the use of an eye-tracker device to evaluate pupillometry (DiNuzzo et al., 2019; Murphy et al., 2014; Schneider et al., 2016; Yellin et al., 2015). The mean pupil diameter decreases during drowsiness (Lowenstein et al., 1963; McLaren et al., 1992; Oken et al., 2006), due to low tonic firing rates of the locus coeruleus neurons and dominant parasympathetic nerve system of the pupil (Gilzenrat et al., 2010; Henson and Emuh, 2010).

In this study, we aimed to derive and independently verify vigilance measures obtained from EEG data acquired during rsfMRI. To replicate and extend prior rsfMRI vigilance findings, the present study recorded simultaneous eyes-open rsfMRI, EEG, eye-tracking, and heart rate signals from healthy participants. First, we determined EEG features associated with pupil dimension and validated these features by correlating them with heart rate changes associated with arousal. Then, we investigated the relationship between these validated EEG features and BOLD fMRI signals to illustrate spatial and temporal characteristics of the brain's vigilance response.

## 2. Materials and Methods

### 2.1. Data Collection

This study was conducted at the Laureate Institute for Brain Research with a research protocol approved by the Western Institutional Review Board (IRB). Although 14 healthy subjects participated in this study, data from four volunteers were excluded due to excessive head motion, falling asleep, or closing their eyes during rsfMRI recording. Therefore, data from 10 participants (4 female, age $M = 23.0$ years, range 19-30 years) were analyzed. Participants were recruited from the general community through electronic and print advertisements. They were undergoing medical and psychiatric screening evaluations at the Laureate Institute for Brain Research, including Structural Clinical Interview for *DSM-IV-TR* Axis I Disorders (First et al., 2002). Exclusion criteria included current pregnancy, general MRI exclusions, and any personal major psychiatric or medical disorders. All participants provided written informed consent and received financial compensation for participation. Three 12-minute eyes-open rsfMRI runs were collected from each participant. Prior to each run, participants were instructed to clear their minds, not think about anything in particular, and try to keep their eyes open and fixated on the cross.

A General Electric (GE) Discovery MR750 whole-body 3T MRI scanner (GE Healthcare, Waukesha, Wisconsin, USA) and a standard 8-channel, receive-only head coil array were used for imaging. A single-shot gradient-recalled echoplanar imaging (EPI) sequence with Sensitivity Encoding (SENSE; (Pruessmann et al., 1999)) was used for fMRI acquisition with the following parameters: FOV/slice thickness/slice gap = 240/2.9/0.5 mm, 41 axial slices per volume, 96 × 96 acquisition matrix, repetition time (TR), echo time (TE) TR/TE= 2000/30 ms, acceleration factor R = 2, flip angle = 90°, sampling bandwidth = 250 kHz. To allow the fMRI signal to reach a steady-state, three EPI volumes (6 s) were excluded from data analysis. For each of three rsfMRI runs, simultaneous physiological pulse oximetry and respiration waveforms were collected (with 50 Hz sampling, using a photoplethysmograph with an infra-red emitter placed under the pad of the subject's left index finger and a pneumatic respiration belt, respectively). To provide an anatomical reference for the fMRI analysis, a T1-weighted magnetization-prepared rapid gradient-echo (MPRAGE) sequence with SENSE was collected with the following parameters: scan time=4 min 58 sec, FOV=240 mm, axial slices per slab=128, slice thickness=1.2 mm, image matrix=256×256, TR/TE=5/1.9 ms, acceleration factor R=2, flip angle=10°, delay time TD=1400 ms, inversion time

TI=725 ms, sampling bandwidth=31.2 kHz. EEG signals were recorded simultaneously with fMRI via a 32-channel MR-compatible EEG system from Brain Products GmbH. The EEG cap consisted of 32 channels, including references, arranged according to the international 10-20 system. One electrode was placed on the subject's back for recording the electrocardiogram (ECG) signal. A Brain Products' SyncBox device was used to synchronize the EEG system clock with the 10 MHz MRI scanner clock. EEG acquisition temporal resolution was 0.2 ms (i.e., 16-bit 5 kS/s sampling), and measurement resolution was 0.1 µV. EEG signals were hardware-filtered throughout the acquisition in the frequency band between 0.016 Hz and 250 Hz. Pupil size was recorded in arbitrary units at a sampling rate of 250 samp/sec using an MRC eye-tracker system (MRC Systems GmbH, Heidelberg, Germany). Transmission Control Protocol/Internet Protocol (TCP/IP) connection was used to synchronize the eye tracker device with the MRI machine clock using MRC System Dynamic-link library (DLL). The pupil size was recorded from the left eye. An example of the eye tracker interface and a 30-sec pupil size recording are provided on Supplementary Figure S1.

## 2.2. Data Analysis

Each of the EEG, fMRI, and eye-tracker modalities require specific preprocessing to reduce noise and artifacts, as well as to recover missing data. After applying preprocessing steps separately for the data within each modality, we combined the clean data from those three modalities. Figure 1 shows a summary of data analysis steps as a function of modality.

### 2.2.1 Preprocessing

EEG is highly sensitive to noise and artifacts. In order to reduce the artifacts and preprocess EEG data, we adopted the pipeline described by (Mayeli et al., 2016) using BrainVision Analyzer 2 software (Brain Products GmbH, Munich, Germany). Imaging artifacts were reduced using the average artifact subtraction (AAS) method (Allen et al., 2000), and EEG signals were down-sampled to 250 Hz. Next, a zero-phase Shift Butterworth fourth order (i.e., slope of 24 dB/octave) band-rejection filters (1 Hz bandwidth) were used to remove fMRI slice selection fundamental frequency (20.5 Hz) and its harmonics, vibration noise (26 Hz), and alternating current (AC) power line noise (60 Hz). Then, a bandpass filter from 0.1 to 80 Hz (Zero-phase Shift Butterworth Filters, slope of 48 dB/octave) was applied to remove signals unrelated to brain activity. Ballistocardiogram (BCG) artifacts were also removed using the AAS (Allen et al., 1998). The cardiac cycles were first automatically detected using the back-ECG electrode via BrainVision Analyzer 2 software and then were manually inspected and adjusted, if required. For each channel, the BCG artifact template was generated using a moving average over 21 cardiac periods and then subtracted from the data. After reducing BCG artifacts, the Infomax independent component analysis (ICA) was utilized for independent component decomposition (Bell and Sejnowski, 1995) over the entire data length with exclusion of intervals that were excessive motion-affected. ICA was applied to the data from 31 EEG electrodes and yielded on 31 independent components (ICs). The topographic map, power spectrum density, time course signal and energy of ICs were used for detecting and removing artifactual ICs. For instance, ocular artifacts are identified by strong spatial projection in the prefrontal area, either by exhibiting high loadings at the most anterior regions for eye blinks or showing as an anterior dipole for saccades, as well as high energy and low-frequency peaks (between 0.5 and 3 Hz) in the frequency domain. The single channel ICs were identified using their topographic maps. That kind of artifact affects one or two channels without showing any effects on other channels, either because that channel has poor contact during recording or it is one of the channels (e.g., T7, T8, TP9 and TP10) that is more sensitive to jaw and head movement (Zotev et al., 2018). Residual BCG artifacts normally show in a bipolar topographic map with frequency activity in Theta band. Finally, muscle artifacts can be identified by high frequency activity and peaks in the Gamma band (30-60 Hz range). The topographic map examples of each type of artifacts as well as neural activity are shown in supplementary Figure S2. After selecting the artifactual ICs for removal, the EEG signal was reconstructed using inverse ICA after selecting ICs related to neural activities.

Pupil size vector includes both missing data due to eye-blinks and noise. In preprocessing the pupil size signal, the "fillmissing" command in MATLAB (MathWorks Inc, Natick, Massachusetts, USA) was used first to interpolate missing numeric data. More specifically, a moving median window with a length of twice the largest gap was used. Then, a zero Phase Butterworth band-pass filter (0.01–0.1 Hz) was applied to correct for very slow drifts and high-frequency oscillations, as suggested in previous works (Schneider et al., 2016; Yellin et al., 2015). The data from subjects with more than one-third of missing data were excluded, resulting in 21 runs from 10 subjects (the percentage of missing data from each run and participant is shown in Table S1 in the Supplementary Material).

Imaging analyses were carried out using the Analysis of Functional NeuroImages software (AFNI, http://afni.nimh.nih.gov/afni/) (Cox, 1996). The afni_proc.py command was used to preprocess the data using the default parameters. The first three volumes were omitted from the analysis to allow the fMRI signal to reach a steady state. The despike option was adopted to replace outlier time points with interpolation. RETROICOR (Glover et al., 2000) and respiration volume per time (RVT) correction (Birn et al., 2008) were applied to remove cardiac- and respiration-induced noise in the BOLD signal. Slice-timing differences were adjusted by aligning to the first slice, and motion correction was applied by aligning all functional volumes to the volume with low-motion determined by the data. EPI volumes were acquired using the 3dvolreg AFNI program with two-pass registration. The volume with the minimum outlier fraction of the short EPI dataset acquired immediately after the high-resolution anatomical (MPRAGE) brain image was used as the registration base. Linear warping was applied to the MNI space and resampled to 2 mm3 voxels. Data were spatially smoothed (6 mm FWHM) and scaled to have a mean of 100 and range of [0-200]. The outcome was then used as an input for GLM analysis along with the regressor of interest. To control for the nuisance variables, we used the 12 motion parameters (3 shift and 3 rotation parameters with their temporal derivatives) and three principal components of the ventricle signal from the signal time course as covariates in the GLM. FreeSurfer 5.3 (http://surfer.nmr.mgh.harvard.edu/) was used to extract white matter and ventricle masks from the anatomical image of an individual subject and then warped them to the normalized fMRI image space. In order to investigate the effects of physiological noise correction (i.e., RETROICOR and RVT), we repeated the analysis without including those corrections in the preprocessing step.

The heart rate for each individual run and subject was computed using a custom MATLAB script by dividing 60 with the average interval (in seconds) between two R-peaks of the ECG. The values were averaged every 4 seconds.

**2.2.2 Postprocessing**

After removing EEG artifact, the following features were extracted using a recent open-source EEG feature extraction software (Toole and Boylan, 2017) in MATLAB from the EEG data (channels F3, F4, Fz, O1, O2, and Oz) from each subject and each run: (1) power spectral density in alpha band (alpha power); (2) power spectral density in beta band (beta power); and (3) the ratio between alpha power and the power in the combined delta and theta bands (alpha ratio; as suggested in literature for the vigilance index during an eyes-closed resting state). We selected these six EEG channels because previous research has shown an association between those channels' features and vigilance (Knaut et al., 2019; Olbrich et al., 2009), and also because these channels are less vulnerable to artifacts, especially muscle artifacts, than others across the scalp.

PSD was estimated via the periodogram method as follows:

$$\hat{P}(f) = \frac{\Delta t}{N} \left| \sum_{n=0}^{N-1} x_n e^{-j2\pi f n} \right|^2, 0 < f < \frac{1}{2\Delta t}$$

where $x_n$ is the EEG time course, and $\Delta t$ is the sampling interval. After calculating the PSD in each channel, the power in each frequency band was averaged among the six channels of interest.

The value for each of the three EEG features was averaged every 4 seconds among the selected channels. Due to the slow pupillary response (Wierda et al., 2012), as well as the missing pupil size data, we selected the 4-sec window to estimate the vigilance level from the eye-tracker. Therefore, averaged pupil size was calculated every 4 seconds as well, after the aforementioned preprocessing to generate the pupil size vector for all further analysis. Pupil size vector was used as the vigilance stage index (i.e., larger pupil size indicates higher vigilance level). The distribution of pupil size vector and selected EEG features showed a statistically significant deviation from normality (*p < 0.05*) using the Shapiro-Wilk test. Hence, the nonparametric Spearman's rank correlation coefficient $\rho$ was used to measure the correlations between EEG features and pupil size across time within each run and each subject. Furthermore, we used a one-sample t-test on the Fisher z-transformed correlation coefficients to investigate the overall positive EEG feature-pupil size correlation among subjects and calculate the effect size. In addition, for each run, after finding the EEG feature mostly associated with pupil size, we investigated the correlation between that EEG feature and heart rate. Again, we used a one-sample t-test on the Fisher z-transformed correlation coefficients to examine the overall positive correlation between those features among subjects (the runs were average for each subject for this analysis) and computed the effect size. The same analysis was carried out with the features extracted from all channels (instead of using only 6 channels) to confirm the reliability of the EEG feature we selected as the vigilance index. Next, pupil size and EEG features were applied to fMRI data analysis as separate regressors. Each of these regressors was convolved with a hemodynamic response function (HRF) and downsampled to 0.5 Hz (to match the TR of 2 s) if needed. Supplementary Figure S3 represents the steps for the fMRI analysis.

## 3. Results

Correlations between EEG features from channels F3, F4, Fz, O1, O2, and Oz (i.e., alpha power, beta power, and the alpha ratio) and pupil size are summarized in Figure 2. Supplementary Figure S4 illustrates the linear regression plots for each participant and each run. We repeated the correlation analyses by extracting EEG features from all channels, and the results are shown in Supplementary Figure S5. The statistical details of correlations between pupil size and EEG features, as well as one sample t-test and effect size on the Fisher z-transformed correlation coefficients results, are shown in Table 1.

The correlation coefficients in Table 1 show an overall positive correlation between pupil size vector and both alpha power and beta power, which indicates higher alpha/beta powers represent larger pupil size (and vigilance level). We further compared the Fisher z-transformed correlation coefficients between alpha power and pupil size, and beta power and pupil size from the selected six EEG channels. The results are as follow: $t(18)= -2.162$, $p= 0.058$. The difference between these correlations shows a trend to a significant difference. Also, the results presented in Figure 2 show the correlation between pupil size and beta power was not significant for 3 out of 21 runs; however, 9 out of 21 runs were not significant for alpha power association with pupil size. Therefore, we used frontal and occipital beta power (FOBP; as the vigilance level index) as a regressor for the fMRI analysis. Figure 3 illustrates the raincloud plots comparing the three sets of correlations using 6 EEG channels, as well as 31 channels. The confidence interval for the difference of means of the r to z transformed correlation coefficients of pupil size and alpha and beta powers using 1000 samples bootstrapping are 95% CI[-0.032,0.183] and 95% CI[-0.040,0.186], for having 6 EEG channels and 31 channels, respectively. The bootstrap confidence interval for a difference in mean was done using a freely available online software called StatKey (Morgan et al., 2014). Figure 4 depicts the correlation map between the BOLD signal with correction for physiological noise and FOBP. The analysis was performed for each voxel, and the statistical map was thresholded with voxel-wise *p < 0.005* and cluster-size corrected *p < 0.05*. The cluster-size threshold was evaluated with AFNI's 3dClustSim using an improved spatial autocorrelation function (ACF; (Cox et al., 2017)); a minimum cluster size of 146 voxels was required to have a corrected *p ≤ 0.05* while using 2-sided third nearest neighbor clustering (NN3). Table 2 illustrates that higher values of beta power were associated with greater BOLD signal in the precentral gyrus, postcentral gyrus, and insular cortex, as well as temporal gyrus and inferior parietal lobule. Supplementary Figure S6 represents the results without any cluster-size threshold. The results related to the correlation map between BOLD signal without including correction for physiological noise and FOBP are shown in Supplementary Figure S7 and Table S2.

Figure 5 illustrates correlations between heart rate and FOBP (one-sample t-test on the Fisher z-transformed correlation coefficients results: *t(9)= 4.625, p=0.001, d=1.462*). Supplementary Figure S8 illustrates the correlation between heart rate and beta power including all EEG channels (*t(9)= 4.797, p= 9.8e-04, d= 1.517*).

Finally, Figure 6 shows the correlation maps between the BOLD signal with correction for physiological noise and pupil size. Supplementary Figure S9 represents the results without any cluster-size threshold. Table 3 shows the details of the brain regions associated with that regressor, which represents a higher pupil size linked to lower fronto-occipital BOLD signal. The results for the same analysis without correction for physiological noise are presented in Supplementary Figure S10 and Table S3.

## 4. Discussion

This study aimed to determine whether EEG features could be used as objective markers of vigilance in healthy human subjects during eyes-open rsfMRI experiments. Three main findings regarding this investigation are as follow: First, frontal (F3, F4, and Fz) and occipital (O1, O2, and Oz) beta power (i.e., FOBP) showed the highest correlation with pupil size; Second, FOBP correlated with several brain regions that have been implicated in modulating vigilance; Third, FOBP also was positively correlated with heart rate. Taken together, these findings support the conclusion that FOBP is an objective and robust measure of vigilance in healthy human subjects.

As shown in Figure 2 and Table 1, FOBP has the highest correlation with the pupil size among different EEG features. An earlier simultaneous EEG-fMRI study (Laufs et al., 2003), suggested that alpha rhythm signal was associated with "inattention" during rest, while beta rhythms were posited to index spontaneous cognitive operations during conscious rest. Another study is required to generate a template for vigilance estimation using this metric to produce an fMRI-based template for estimating the subject's vigilance. For the rest of our analysis, we considered FOBP as the vigilance index during our eyes-open resting state.

Although we tested the correlation between the ratio of the PSD in the alpha band and the power in the combined delta and theta band (the most common measure of vigilance during eyes-closed resting state) and pupil size, that association was not significant, which it could indicate the difference between EEG vigilance measures in eyes-open and eyes-closed conditions. The variability between findings in eyes-closed and eyes-open EEG vigilance measures was reported in previous studies. For instance, (Barry et al., 2007) found a significant correlation between mean alpha level across all channels and skin conductance levels as an index for vigilance across the eyes-closed condition. However, no correlations between skin conductance levels and alpha power were reported in that study during their eyes-open condition, so it is possible that the alpha ratio-arousal association only holds for eyes-closed data. A very recent simultaneous EEG-fMRI study (Falahpour et al., 2018) used the ratio of power in the alpha band over the power in delta and theta band as the vigilance index for both eyes-open and eyes-closed conditions. Their results suggested a significant difference between vigilance index BOLD correlation map in thalamus and DMN among eyes-open and eyes-closed conditions (Falahpour et al., 2018). As it was noted in (Chang et al., 2016), one of the reasons for lower association between alpha power/ratio and vigilance is the presence of unconstrained eye open/closure, which it could interfere with the interpretation of alpha power fluctuations while we asked the subjects to try to keep their eyes open. Here, with utilizing the eye-tracker for measuring the subject's vigilance, we showed that beta power is a better index compared to alpha power for the vigilance level during eyes-open resting state.

After finding EEG features with the highest correlation with pupil size, that feature served as a regressor for fMRI analysis, as illustrated in Figure 4. The correlation map for FOBP shows positive correlations with several brain regions, including the bilateral insula, inferior parietal lobule, and supplementary motor area. Interestingly, these brain regions have been reported in previous studies investigating the correlation maps of EEG vigilance indices during eyes-closed rsfMRI (Falahpour et al., 2018; Olbrich et al., 2009; Sadaghiani et al., 2010). Further, these brain regions are all parts of the anti-correlated network (ACN) (De Havas et al., 2012; Fox et al., 2005), which have shown anti-correlated activation with the DMN both during the rsfMRI and task.

A previous study (Yuan et al., 2013) showed that using RVT and RETROICOR for physiological correction has fewer effects on EEG (alpha power) correlation map during eyes-open compared to eyes-closed rsfMRI. To investigate the effects of physiological noise correction, we showed the FOBP and pupil correlation maps in the supplement without including physiological noise correction. The results presented in Tables 2, and S2 show that the same brain regions were found correlated with FOBP regardless of physiological noise correction. It is worth noting, however, that without using physiological noise correction, the BOLD signal in the cerebellum (e.g., declive and culmen) are correlated with FOBP. This is in line with the results presented in (Yuan et al., 2013), which found the RETRICOR and RVT correction has a minimum effect on the EEG correlation map with the BOLD signal during eyes-open rsfMRI. Also, the BOLD activity (after physiological noise correction) from several brain regions that was found correlated with pupil size was repeated when we do not use the noise correction.

Furthermore, to validate our findings, we looked at the correlation between the selected EEG vigilance features and heart rate. As previous studies, such as (Olbrich et al., 2009) showed, higher vigilance was associated with a higher heart rate. Figure 5 represents such a positive correlation. Earlier, we found a positive correlation between pupil size and beta power, and that feature has a positive correlation with heart rate. Therefore, heart rates increased with a higher vigilance level.

Additionally, to further investigate the association between EEG beta power and vigilance, we repeated the analysis with the extraction of EEG features from all 31 channels (instead of 6 channels). As Table 1 and Supplementary Figure S5 show, the association between pupil size and both alpha and beta powers, as well as correlation between beta power and heart rate, are still significantly higher zero using all 31 EEG channels. However, the results from six and all 31 channels are slightly more varied for EEG features related to the beta band as compared to the alpha band (e.g., five runs out of 21 do not show significant correlation between pupil size and beta power for 31 channels, compared to three runs for six channels), and that could be because of the residual muscle and imaging artifacts in the beta band, especially in edge electrodes (e.g., TP9 and TP10). We specifically selected those six channels first because they have been used in previous studies for investigating the vigilance level (Hegerl et al., 2008; Olbrich et al., 2009) and second because they are less vulnerable to noise and artifacts.

Finally, pupil size served as a regressor for fMRI. We found negative correlations between activity between pupil size and visual and sensorimotor cortices of the brain, which is in a great agreement with previous simultaneously eye-tracker-fMRI studies (DiNuzzo et al., 2019; Murphy et al., 2014; Schneider et al., 2016; Yellin et al., 2015). However, there were some variabilities between previous studies regarding the brain regions positively correlated with pupil size. For instance, (Yellin et al., 2015) found a positive correlation between pupil size and DMN, and (Schneider et al., 2016) found thalamus, caudate nucleus, and cerebellum are positively associated with pupil size. Those associations between the BOLD signal and pupil dimension were replicated neither in other studies nor in ours.

It should be noted that in 3 runs out of 21 (i.e., ~14%), we did not see a significant correlation between pupil size and FOBP. Such variability was expected due to two possible reasons. First, the low vigilance fluctuation within a given scan can reduce the brain activity modulation (Falahpour et al., 2018) and changing the EEG power. Therefore, we might see less association between pupil size and EEG beta power. For instance, for one of the runs that we did not observe a significant correlation between FOBP and pupil size (Subject 7 Rest1), the missing pupil size percentage is less than 4%, which could indicate the low vigilance fluctuation. Another possible explanation for this variability is that, although we carefully preprocessed EEG and eye-tracker data, however, there might be still some residual artifact/noise on each of those datasets.

Furthermore, our initial goal was to investigate the subjects' vigilance variability among different runs. However, we could use all three rsfMRI runs for only three participants, due to missing pupil size information for some of the resting-state runs for the other individuals. Therefore, we did not have enough statistical power to provide such an investigation.

It should be pointed out that, although an eye-tracker was used as an indirect measure of vigilance during eyes-open resting state, there is no guarantee that participants will keep their eyes open during the entire scan time. Therefore, it will be necessary to test if FOBP could serve as an appropriate EEG vigilance measure during the eyes-closed resting state. Since an eye-tracker device is not a helpful tool for eyes-closed scans. Instead, skin conductivity and heart-rate parameters are suggested as possible indicators.

## 5. Conclusion

Simultaneous EEG-fMRI-eye-tracker experiments have been suggested in an effort to determine, verify, and measure participant arousal/vigilance level during rsfMRI. However, such an experimental setup requires specific, and often costly, hardware and software. Furthermore, the analysis of data within each of these modalities requires an expert workforce. Therefore, running this experiment is not feasible for most research centers. In this study, we designed the first multimodal EEG-rsfMRI-eye-tracker experiment in human participants (to the best of our knowledge) to find EEG vigilance metrics during eyes open resting-state fMRI. Our results revealed an association between frontal and occipital beta power and degree of vigilance during an eyes-open resting state. This EEG measure could be easier to determine during simultaneous EEG-fMRI and even provide a real-time measure of subject vigilance in resting-state fMRI. We validated the results using simultaneously-recorded heart rate and fMRI. Interestingly, independently-verified heart rate changes can also provide an easy-to-determine measure of vigilance during rsfMRI.

**Acknowledgments**

This work has been supported by Laureate Institute for Brain Research, the William K. Warren Foundation, and by National Institute of General Medical Sciences, National Institutes of Health Award 1P20GM121312. The funding sources have no influence in study design, the data collection, analysis, data interpretations, in the writing of the manuscript, and the decision to submit the article for publication.

**Availability of data**

The data that support the findings of this study are available from the corresponding author upon reasonable request.

**Declaration of interest**

The authors declare no competing financial interests.

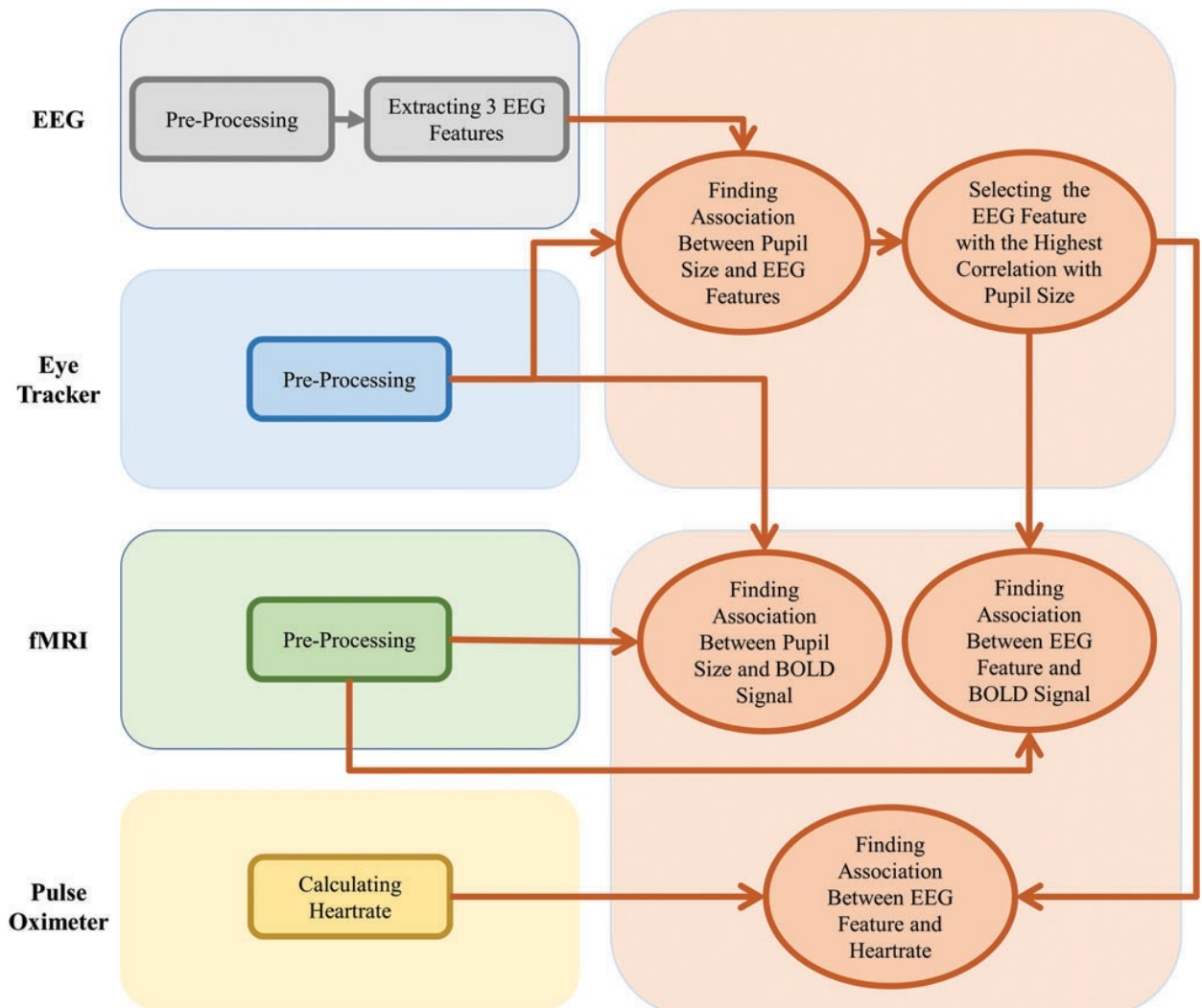

FIG. 1. The data analysis flowchart. Each of the modalities collected in this study required its own preprocessing steps to remove noise and artifacts and to recover missing data. After preprocessing EEG data, three features were extracted from the channels F3, F4, Fz, O1, O2, and Oz EEG data as follows: (1) power spectral density in the alpha band, (2) power spectral density in the beta band, and (3) alpha power ratio. The associations between these features and pupil size were investigated, and the feature with the highest association was selected for inclusion into further analysis. Next, the associations between that feature and the BOLD signal, as well as heart rate, were investigated. The pupil size was used as a regressor to evaluate the relationship with the BOLD signal. We also investigated the association between those three EEG features extracted from all 31 channels and pupil size and heart rate. BOLD, blood-oxygen-level-dependent; EEG, electroencephalography. Color images are available online.

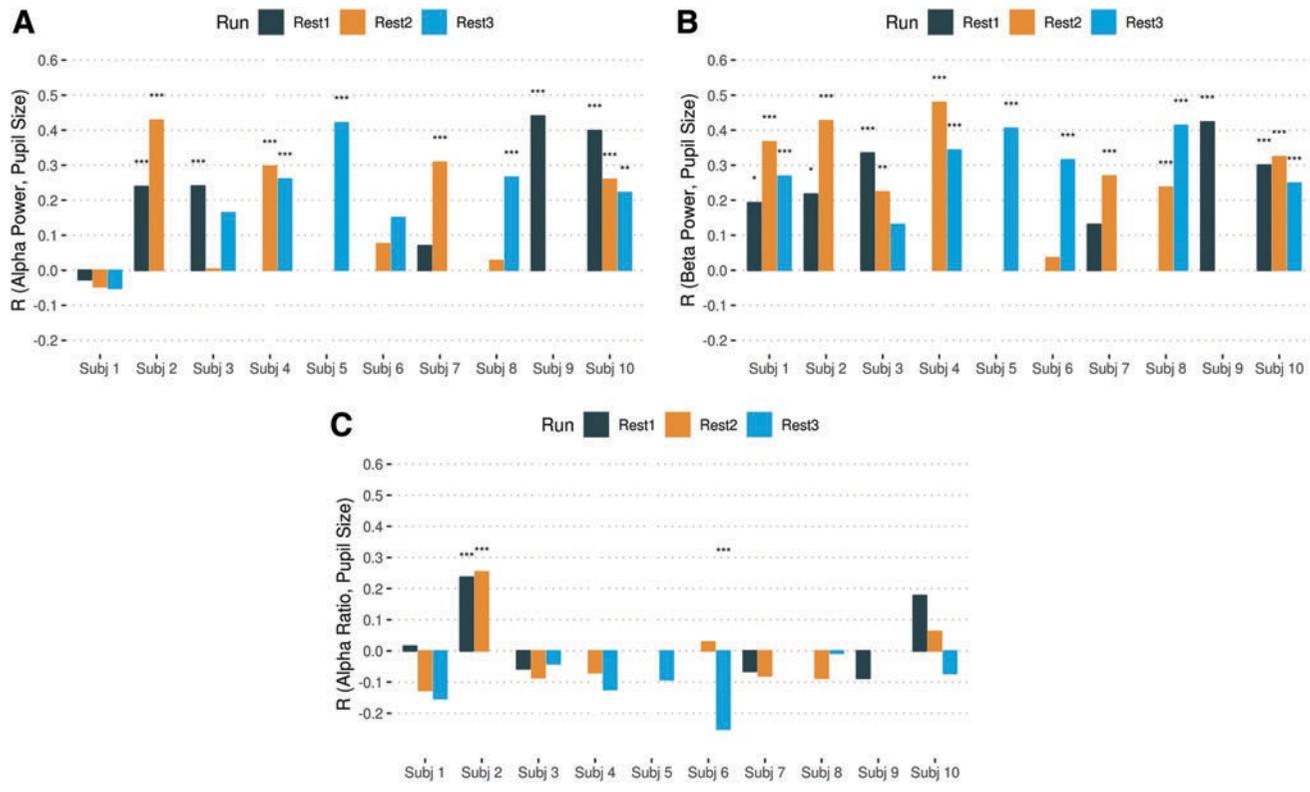

FIG. 2. Association between EEG features and pupil size. The correlation between EEG features from frontal (F3, F4, Fz) and occipital (O1, O2, Oz) channels, that is, (A) power spectral density in the alpha band, (B) power spectral density in the beta band (FOBP), and (C) alpha power ratio and pupil size are shown for each run and each subject. The runs with missing data are left empty. The asterisks show significant levels of correlation (corrected for multiple comparisons). FOBP, frontal and occipital beta power. Color images are available online.

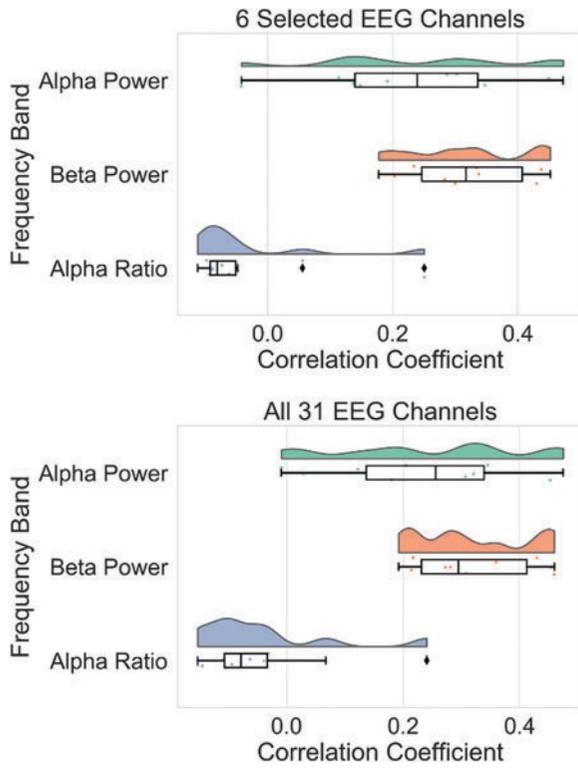

FIG. 3. The raincloud plots comparing the three sets of correlations between pupil size and EEG features. The top figure shows the raincloud plot for the *r*-to-*z* transformed cor- relation between pupil size and EEG features using six EEG channels (the runs were average for each subject for this analysis). The bottom plot shows the same analysis using all 31 EEG channels. Color images are available online.

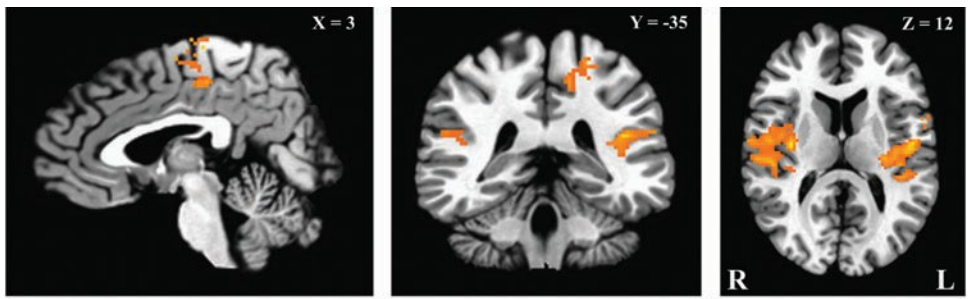

FIG. 4. Power spectral density in FOBP correlation map. The FOBP was used as a regressor in fMRI GLM analysis. The figure shows the cluster survived with sampling at uncorrected $p < 0.005$ and at cluster-size thresholded with AFNI's 3dClust- Sim using an improved spatial ACF, with a minimum cluster size of 146. ACF, autocorrelation function; AFNI, Analysis of Functional NeuroImages; fMRI, functional magnetic resonance imaging; GLM, general linear model. Color images are avail- able online.

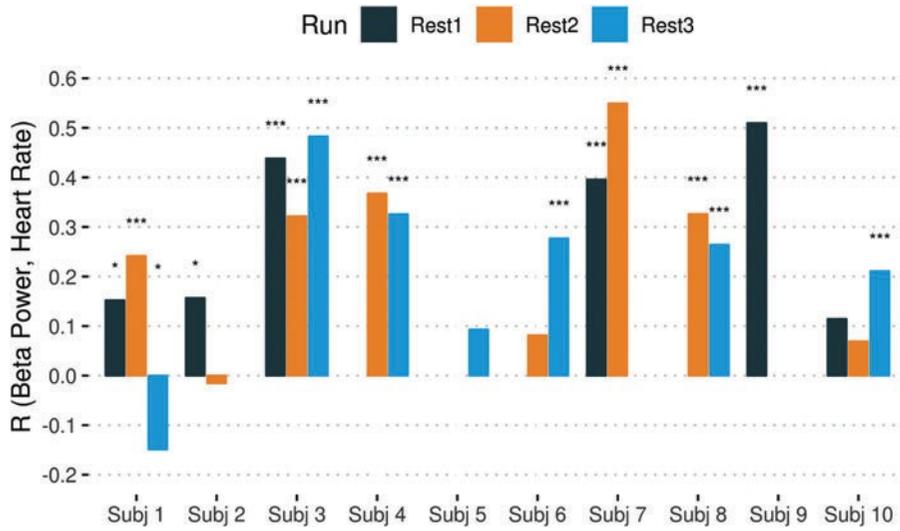

FIG. 5. Association between FOBP and heart rate. The correlations between FOBP and heart rate are shown for each run and each subject. The runs with missing data are left empty. The asterisks show significant levels of the correlation. Color images are available online.

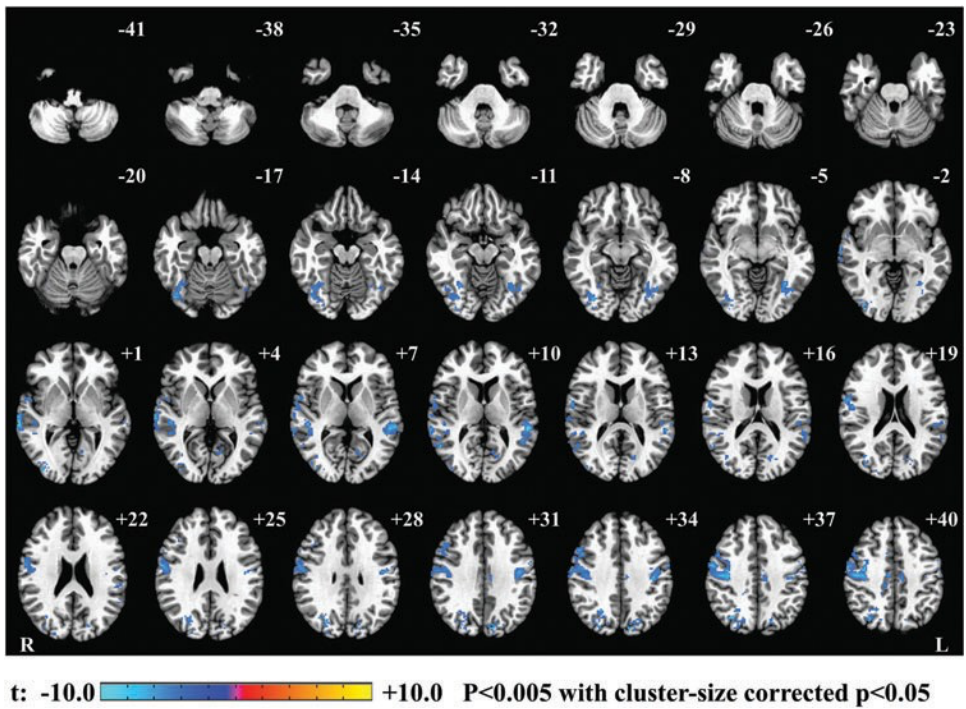

FIG. 6. Pupil size correla- tion map. Pupil size was used as a regressor in fMRI GLM analysis. The figure shows the clusters survived with sampling at uncorrected p < 0.005 and at cluster-size thresholded with AFNI's 3dClustSim using an im- proved spatial ACF, with a minimum cluster size of 148. Color images are available online.

Table 1. The Details of Correlations Between Pupil Size and Electroencephalography Features (Extracted from 6 Selected Channels and All 31 Channels) Among Subjects (for Subjects with More than One Run, the Correlation Coefficients Were Averaged Among Runs, Resulting in One Value for Each Subject)

|  | Average of correlation coefficient | STD of correlation coefficient | One-pair t-test on r to z transformed | | | Cohen's effect size |
|---|---|---|---|---|---|---|
|  |  |  | df | p | t-stat |  |
| Six selected channels |  |  |  |  |  |  |
| Alpha power | 0.231 | 0.115 | 9 | 0.001 | 4.739 | 1.499 |
| beta Power | 0.306 | 0.107 | 9 | 2.90E-06 | 10.254 | 3.241 |
| alpha ratio | −0.036 | 0.109 | 9 | 0.338 | −1.01 | −0.319 |
| All 31 channels |  |  |  |  |  |  |
| Alpha power | 0.233 | 0.155 | 9 | 0.001 | 4.627 | 1.463 |
| Beta power | 0.306 | 0.091 | 9 | 3.80E-06 | 9.922 | 3.136 |
| Alpha ratio | −0.043 | 0.117 | 9 | 0.283 | −1.143 | −0.361 |

The *t*-test parameters and Cohen's effect size were calculated on the Fisher *z*-transformed correlation coefficients.

Table 2. Brain Regions Correlated to Power Spectral Density in Frontal and Occipital Beta Power

| | Cluster | | | | Cluster size (No. of voxels) |
|---|---|---|---|---|---|
| | Peak coordinates (Talairach) | | | | |
| | x | Y | z | t-Score | |
| Insula (R), claustrum (R), superior temporal gyrus (R), precentral gyrus (R), transverse temporal gyrus (R), postcentral gyrus (R) | 33 | −13 | 12 | 9.416 | 1052 |
| Insula (L), superior temporal gyrus (L), claustrum (L), precentral gyrus (L), postcentral gyrus (L), inferior parietal lobule (L) | −51 | −9 | 18 | 9.719 | 792 |
| Medial frontal gyrus (R), paracentral lobule (R), paracentral lobule (L) | 3 | −23 | 64 | 8.004 | 191 |
| Paracentral gyrus (L), postcentral gyrus (L), inferior parietal lobule (L) | −23 | −39 | 54 | 6.924 | 155 |

Table 3. Brain Regions Correlated to Pupil Size

| | Cluster | | | | |
|---|---|---|---|---|---|
| | Peak coordinates (Talairach) | | | | Cluster size (No. of voxels) |
| | X | y | z | t-Score | |
| Middle frontal gyrus (R), precentral gyrus (R), postcentral gyrus (R), middle cingulate gyrus, paracentral gyrus, medial frontal gyrus | 25 | −27 | 56 | −12.903 | 3494 |
| Declive (R), fusiform gyrus (R), inferior occipital gyrus (R), middle occipital gyrus (R), precuneus (R), lingual gyrus (R) | 35 | −85 | −2 | −10.625 | 684 |
| Superior temporal gyrus (L), postcentral gyrus (L) | −53 | −35 | 10 | −13.792 | 299 |
| Precentral gyrus (L) | −43 | −13 | 52 | −9.921 | 262 |
| Parahippocampal gyrus (L), fusiform gyrus (L), inferior occipital gyrus (L), middle occipital gyrus (L), declive (L) | −33 | −55 | −4 | −7.54 | 205 |